\documentclass[conference]{IEEEtran}
\IEEEoverridecommandlockouts
\usepackage{cite}
\usepackage{amsmath,amssymb,amsfonts}
\usepackage{algorithmic}
\usepackage{graphicx}
\usepackage{dcolumn}
\usepackage{bm}
\usepackage{simpler-wick}
\usepackage{booktabs}
\usepackage{amsmath}
\usepackage{physics}
\usepackage{nicefrac}
\usepackage{float}
\usepackage{multirow}
\usepackage{tikz}
\usepackage{dsfont}
\usepackage{adjustbox}
\usepackage{hyperref}
\usetikzlibrary{quantikz2}
\usepackage[normalem]{ulem}

\newcommand{\Pf}{\mathrm{Pf}}
\usepackage{textcomp}
\usepackage{xcolor}
\def\BibTeX{{\rm B\kern-.05em{\sc i\kern-.025em b}\kern-.08em
    T\kern-.1667em\lower.7ex\hbox{E}\kern-.125emX}}
\begin{document}

\title{Fermionic Machine Learning\\
\thanks{Supported by the Minist\`{e}re de l'\'{E}conomie, de l'Innovation et de l'Énergie du Qu\'{e}bec through its Research Chair in Quantum Computing, an NSERC Discovery grant. JG acknowledges support from the QSciTech CREATE-NSERC program. AS acknowledges support from the Canada First Research Excellence Fund through an Institut quantique postdoctoral fellowship. This work made use of compute resources by Calcul Québec and the Digital Research Alliance of Canada.}
}

\author{\IEEEauthorblockN{J\'{e}r\'{e}mie Gince}
\IEEEauthorblockA{\textit{Département de physique and Institut quantique} \\
\textit{Université de Sherbrooke,}\\
Sherbrooke,  QC, J1K 2R1, Canada \\
}
\and
\IEEEauthorblockN{Jean-Michel Pag\'{e}}
\IEEEauthorblockA{\textit{Département de mathematique} \\
\textit{Université de Sherbrooke,}\\
Sherbrooke,  QC, J1K 2R1, Canada \\
}
\and
\IEEEauthorblockN{Marco Armenta}
\IEEEauthorblockA{\textit{AlgoLab, Institut quantique} \\
\textit{Université de Sherbrooke,}\\
Sherbrooke, QC, J1K 2R1, Canada  \\
}
\and
\IEEEauthorblockN{Ayana Sarkar}
\IEEEauthorblockA{\textit{Département de physique and Institut quantique} \\
\textit{Université de Sherbrooke,}\\
Sherbrooke,  QC, J1K 2R1, Canada }
\and
\IEEEauthorblockN{Stefanos Kourtis}
\IEEEauthorblockA{\textit{Département de physique and Institut quantique} \\
\textit{Université de Sherbrooke,}\\
Sherbrooke,  QC, J1K 2R1, Canada \\
}
}

\maketitle

\begin{abstract}
We introduce \emph{fermionic machine learning} (FermiML), a machine learning framework based on fermionic quantum computation. FermiML models are expressed in terms of parameterized matchgate circuits, a restricted class of quantum circuits that map exactly to systems of free Majorana fermions. The FermiML framework allows for building ``fermionic'' counterparts of any quantum machine learning (QML) model based on parameterized quantum circuits, including models that produce highly entangled quantum states. Importantly, matchgate circuits are efficiently simulable classically, thus rendering FermiML a flexible framework for utility benchmarks of QML methods on large real-world datasets. We initiate the exploration of FermiML by benchmarking it against unrestricted PQCs in the context of classification with random quantum kernels. Through experiments on standard datasets (Digits and Wisconsin Breast Cancer), we demonstrate that FermiML kernels are on-par with unrestricted PQC kernels in classification tasks using support-vector machines. Furthermore, we find that FermiML kernels outperform their unrestricted candidates on multi-class classification, including on datasets with several tens of relevant features. We thus show how FermiML enables us to explore regimes previously inaccessible to QML methods.
\end{abstract}


\section{Introduction}
\noindent Quantum machine learning (QML) is an interdisciplinary
area of research that explores the potential uses of quantum computing in machine learning (ML)~\cite{LMR2013,PW2014,SSP2015,SP2018,Cili2018,ZJQ2023}. QML
aspires to utilize quantum phenomena (superposition, entanglement, interference) to process data fundamentally differently from classical computers. 
Although QML concepts have been theorized since the
early days of quantum computing~\cite{SG2004,BP1997,VM2000,Schu2003,KAK1995,BJ1998}, recent advances in quantum computing hardware have sharply increased the interest for development of practical QML applications in a short- to mid-term horizon~\cite{PW2014,SSP2015,DW2020,SP2018}.

A hotly debated question has been whether there exist QML approaches that operate on noisy quantum processors
lacking fault tolerance and that can outperform
classical algorithms in specific ML tasks, yielding what
is often termed as quantum advantage. While claims that
the answer to this question may be affirmative have surfaced~\cite{HHL2009,WBL2012,ZNL2021,RML2014,SBSW2020,BLSF2019,wiebe2015,KP2022,Wold2021,ASZLFW2021,LBRquantsup2017,morvan2023phasequantsup}, the highly tailored benchmarks involved
in these proposals cast doubts to the relevance of existing QML techniques in practical settings, with some even questioning the pursuit of quantum advantage as it is currently construed ~\cite{LAT2021, SchuKill2022}.

Near-term QML methods are predominantly based on variational quantum algorithms (VQAs)~\cite{cerezoVQA2021}.  In VQAs, parameterized quantum circuits are tuned via standard classical optimization performed by a classical computer and using probability outcomes of specific quantum circuits to evaluate a task-dependent loss function that is being optimized. Near-term QML methods based on PQCs face two major challenges, namely, barren plateaus and lack of scalability~\cite{MBSBN2018}. 

Barren plateaus are parameter regimes in which gradient-based optimization becomes inefficient due to vanishingly small gradients. It has been observed that barren plateaus are widespread whenever QML is applied beyond the toy-problem scale, thus plaguing most QML approaches~\cite{MBSBN2018,GWOB2019, ACCCC2021,CSVCC2021, QWZGG2023}. While barren plateaus pose an obstacle to the deployment of QML, it is yet unclear whether they are a generic feature of QML based on PQCs rather than a side effect of model choice, with recent work pointing to models in which barren plateaus are provably absent~\cite{cerezo2023,diaz2023}.

On the other hand, demonstrating the performance of
QML protocols upon scaling towards realistic data remains intractable. As far as real-world datasets are concerned, both experiments on near-term QPUs and classical simulation of existing QML models are likely to remain intractable in the foreseeable future. Due to this, the capacity to determine the combinations of quantum resources that can meaningfully boost ML performance in practical settings is severely limited. This lack of scalability remains an obstacle in benchmarking QML even in the absence of barren plateaus. 

With this challenge in mind, we introduce fermionic machine learning (FermiML), a machine learning framework based on fermionic quantum computation. The FermiML framework allows for building key ingredients of quantum mechanics - superposition, interference, and arbitrarily high entanglement - into QML models that are scalable. FermiML models are expressed in terms of parameterized matchgate circuits, a restricted class of quantum circuits that map exactly to systems of free fermions in one dimension and are hence efficiently simulable classically~\cite{JM08matchgate}. FermiML thus offers a toolbox for benchmarking QML protocols on realistic data sets by restricting the corresponding quantum circuits inside the matchgate manifold. On one hand, FermiML establishes a yardstick of performance for any QML algorithm, since any algorithm would need to outperform FermiML to be considered useful. On the other hand, FermiML can be incorporated into classical (deep) learning models, leading to novel quantum-inspired classical algorithms for ML, in the vein of a growing body of classical algorithms that emulate quantum principles to outperform the previous state-of-the-art~\cite{GLT2018,Tang2019,GST2022,ADBL2020,CKLM2023,SBSSBW2023}.

In this work, we begin the exploration of FermiML in the context of quantum kernel learning (QKL)~\cite{schuld2021supervised}. QKL is a straightforward extension of kernel methods in the quantum realm. In QKL, classical data points $x$ are mapped to density matrices through a map $\rho : x \xmapsto{}\rho(x) \in \mathcal{H} = \mathds{C}^{2^{N} \times 2^{N}}$~\cite{KBS2021IB}. The mapping is achieved through evolving an easy-to-prepare quantum state over $N$ qubits, typically $|\mathbf{0}\rangle = |0\rangle ^{\otimes N}$, with a parameterized quantum circuit $U(x)$, so that $\rho(x) = \ketbra{\psi(x)}{\psi(x)}$ with $\ket{\psi(x)} = U(x)|\mathbf{0}\rangle$. The dimension of the feature space is thus determined by the number of qubits. The encoding of $x$ into $\rho(x)$ is determined by the geometry and parameterized gates of the quantum circuit, as we discuss below. The quantum kernel is then defined as $\mathcal{U}(x, x') = \Tr[\rho(x)\rho(x')]$ and measures the similarity between the quantum states associated with $x$ and $x'$~\cite{KBS2021IB}. This kernel is evaluated to a given accuracy by repeated measurements of $\ket{\psi(x)}$.

Once a kernel for a given dataset is obtained, kernel
methods can be employed to perform standard ML tasks,
such as classification or regression~\cite{STC2004}. The parameters of a quantum kernel can either be judiciously pre-determined or variationally optimized to better align the kernel with the data, but the latter strategy potentially gives rise to barren plateaus. Since our goal in this work is to tackle the scalability of QML to real datasets, below we study exclusively predetermined kernels for which the barren plateau issue is irrelevant.

\section{Matchgate circuits and free fermions}

FermiML is based on fermionic quantum computation (FermiQC)~\cite{BraKit2002}, a restricted class of gate-based quantum computations grounded in the physics of free fermion systems~\cite{knill01,Terh2002, bravyi2008contraction,JM08matchgate,jozsa2008embedding,BG2011,Brod16,HJKS2020}. FermiQC contains the set of computations performed by so-called holographic algorithms~\cite{Val01, Val02, Val05, Val08}, originally geared towards the solution of counting problems~\cite{CaiChowHol2007, CaiLu2008Holalg, CaiLu2009Hol,CaiLuX2009Holant, CaiLu2007Hol}. FermiQC is implemented by circuits of matchgates, which are algebraically restricted parity-preserving nearest-neighbour quantum gates. The form of a 2-qubit matchgate is
\begin{equation}
\label{umatchgate}
    U(A,W) =\begin{bmatrix}
             a & 0 & 0 &b\\ 0 & w & x & 0 \\ 0 & y & z & 0\\ c & 0 & 0 & d
             \end{bmatrix} \,,
\end{equation}
where $A = \bigl(\begin{smallmatrix}
a&b\\ c&d
\end{smallmatrix} \bigr)$ and $W = \bigl(\begin{smallmatrix}
w&x\\ y&z
\end{smallmatrix} \bigr)$
satisfy the constraint $\det A = \det W$. The matrices $A$ and $W$ represent the two sub-blocks in even and odd parity subspaces of a 2-qubit Hilbert space in the ordered basis $\{|00\rangle, |01\rangle, |10\rangle, |11\rangle\}$. Circuits built of such gates acting exclusively on pairs of neighboring qubits arranged in a linear array are classically simulable in polynomial time~\cite{JM08matchgate}.

The connection with the physics of free fermions is established by expressing each matchgate $U = e^{iH}$, where $H$ may be expanded as a linear combination of quadratic Majorana monomials~\cite{Terh2002,JM08matchgate},
\begin{align}
\label{gateHamapp}
    H = i \sum\limits_{\mu\neq\nu = 1}^{2N} h_{\mu \nu} c_{\mu}c_{\nu} \,,
\end{align}
where the range of the summation has been extended over all $N$ qubits. The coefficients $h_{\mu \nu }$ are elements of a real, antisymmetric, $2N \times 2N$ matrix. The $c_{\mu}$ are Majorana spinors whose components obey anti-commutation relations: $\{c_{\mu}, c_{\nu}\} = c_{\mu}c_{\nu} + c_{\nu}c_{\mu} = 2 \delta_{\mu \nu} I$, for $\,\,\mu,\nu = 1,..,2N$. The linear span of the Majorana spinors is preserved under conjugation by any matchgate unitary:
\begin{align}
\label{Rmat}
    Uc_{\mu}U^{\dagger} = \sum\limits_{\nu = 1}^{2N} R_{\mu \nu} c_{\nu} \,,
\end{align}
where $R$ is a $2N\times 2N$ matrix determined by reversing the above equation, namely,
\begin{align}
\label{rcomp}
   R_{\mu\nu} &= \frac{1}{4} \Tr{\qty(U c_\mu U^\dagger)c_\nu} \,.
\end{align}
Using $R$, the fermionic kernel can be expressed as,
\begin{align}
\label{outfinal}
     \mathcal{U}(x,x') = \sum\limits_{\mu_{1},\nu_{1},...,\mu_{N},\nu_{N}}^{2N} &T_{j_1,\mu_1} T_{j_1,\nu_1}^*...T_{j_{N},\nu_{N}}^*T_{j_{N},\mu_{N}} \\ \nonumber \times &\left\langle \boldsymbol{0}\left|c_{\mu_1}c_{\nu_1}\cdots c_{\nu_{N}}c_{\mu_{N}}\right|\boldsymbol{0}\right\rangle\,,
\end{align}
where for the j$^{\textrm{th}}$ qubit line, we define
\begin{align}
    \label{Tmat}
    T_{j,\nu} = \frac{1}{2}(R^{\mathsf{T}}_{2j-1,\nu} + iR^{\mathsf{T}}_{2j,\nu})\,.
\end{align} 
The summation in Eq.~\eqref{outfinal} can be performed efficiently (see Appendix).

\section{Data Encoding, Circuit Architecture and Datasets}

\begin{figure}[t]
    \centering
    \includegraphics[width = 8.5cm]{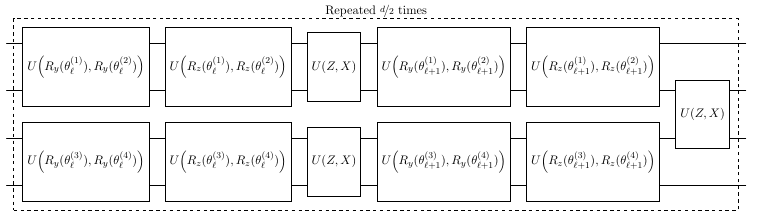}
    \caption{The fPQC has $N$ qubits and $d/2$ layers of parameterized $U\left(R_{y}(\theta_1),R_{y}(\theta_2)\right)$ and $U\left(R_{z}(\theta_1),R_{z}(\theta_2)\right)$ gates followed by entangling gates such as $U(Z,X)$.}
    \label{fig:fpqc}
\end{figure}

As depicted in Figure~\ref{fig:fpqc}, our kernels are constructed through a series of alternating layers of parameterized $U\left(R_{y}(\theta_1),R_{y}(\theta_2)\right)$ and $U\left(R_{z}(\theta_1),R_{z}(\theta_2)\right)$ gates and non-parameterized $U(H,H)$ and $U(Z,X)$ gates, where $U$ comes from Eq.~\eqref{umatchgate}. 
Classical data is loaded into the kernel circuit via the encoding
\begin{equation}
    \theta_j = c_\theta \theta_r + c_x x_j,
\end{equation}
where $\theta_r \in [0, 1]$ is a randomly chosen parameter, $x_j$ corresponds to the $j^{th}$ feature of the data point $x$, and $c_\theta$ and $c_x$ serve as scaling factors, both set to $\nicefrac{\pi}{2}$~\cite{haug_quantum_2023}. The depth of the kernel circuit is
\begin{equation}
    d = \bigg\lceil\frac{\chi}{N}\bigg\rceil,
\end{equation} 
where $\chi$ is the number of features in the dataset.
We compare FermiML kernels with $U(H,H)$ (denoted as hfPQC) and and $U(Z,X)$  (denoted as fPQC) entangling layers against unrestricted kernels (PQC) in terms of classification accuracy. Additionally, we include unentangled fermionic ($\otimes$fPQC) kernels and unentangled unrestricted ($\otimes$PQC) kernels as points of reference. Fermionic circuits are simulated efficiently using our own MatchCake library\footnote{Code available at \href{https://github.com/MatchCake/MatchCake}{https://github.com/MatchCake/MatchCake}.} whereas for generic quantum kernels we use the PennyLane statevector simulator~\cite{bergholm2022pennylane}. We use scikit-learn to train a support vector machine (SVM) on each dataset, employing the kernel function $\mathcal{U}$ to compute the similarities, forming the kernel matrix~\cite{STC2004}, between all data points. Using this kernel matrix, the SVM's parameters are optimized by separating the points in the projected space. For multi-class classification, we use the one-versus-all strategy, where a distinct SVM model is employed to distinguish each class from the others. This method ensures that each class is uniquely represented and classified~\cite{PedScikit2011}.

We use two standard classical datasets as benchmarks for classification with FermiML:  Wisconsin Breast Cancer (WBC) dataset~\cite{misc_breast_cancer_wisconsin_(diagnostic)_17} (569 samples, 30 features, 2 classes), and Digits dataset~\cite{misc_optical_recognition_of_handwritten_digits_80} (1797 samples, 64 features, 10 classes). For all datasets and kernel types, we perform 5-fold cross-validation. Performance comparisons on training and test data for various models corresponding to both WBC and Digits datasets are given in Fig.~\ref{fig:allplotsGA} of the Appendix.

\section{Results and Discussion}

\begin{figure}[!tbp]
    \centering
    \includegraphics[width=0.99\columnwidth]{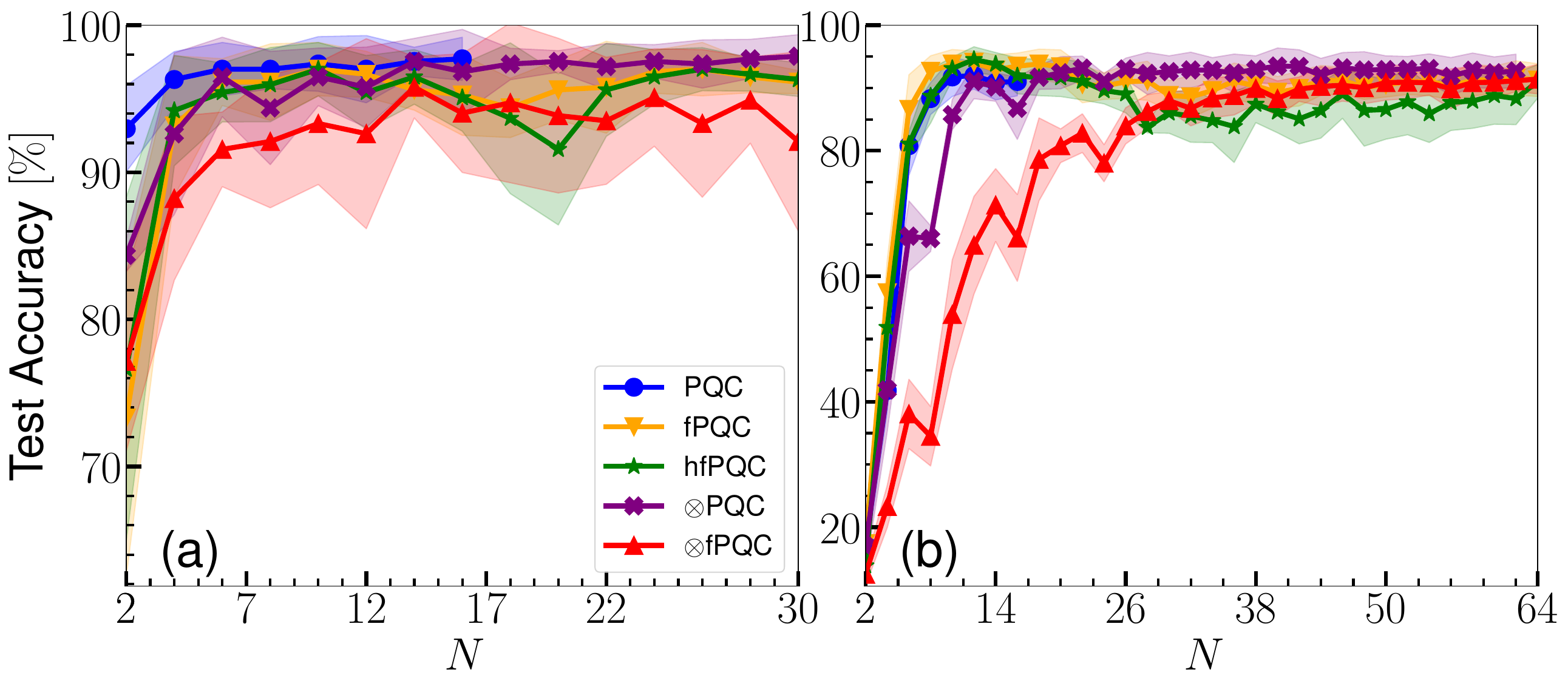}
    \caption{Test accuracies of various models for WBC (a) and Digits (b) datasets as a function of the number of qubits $N$. Shaded regions represent the standard deviation of the accuracy. Note the different axis ranges in the two panels.}
    \label{fullclass}
\end{figure}

For WBC, the plot in Fig.~\ref{fullclass}(a) demonstrates test accuracies for various models against the number of qubits $N$ and indicates accuracies close to 95\%. The unentangled fermionic kernel ($\otimes$fPQC) shows consistently lower accuracies compared to the entangling kernels. While the matchgate models exhibit slightly lower accuracies compared to the PQC with traditional quantum unitaries, simulating PQCs beyond $N = 16$ qubits becomes highly resource-intensive due to the exponential scaling of the statevector. Conversely, simulation runtime scales polynomially for all FermiML models, making it feasible to easily scale to $N = 30$, matching this dataset's feature count and promising an edge of FermiML for QML tasks that require large numbers of qubits. Unentangled unrestricted ($\otimes$PQC) kernels can be simulated efficiently using matrix product state-based tensor network techniques~\cite{Vid2003}, thus allowing for polynomial-time simulation as well. In Figure~\ref{fullclass}(b), the test accuracies of different models on the Digits dataset are presented. Here, both fPQC and hfPQC exhibit slightly better accuracy compared to the PQC. When considering models with fewer qubits, up to about $N = 26$, the performance of $\otimes$fPQC is consistently inferior, indicating that entanglement plays an important role in this regime for the less expressive constrained gate sets that make up the FermiML kernels.  As the number of qubits increases, the accuracies of fPQC, hfPQC, and the unentangled $\otimes$fPQC all converge, meeting within error bars at around $N = 30$ suggesting that the specific details of entangling gates in our case ($U(H,H)$ and $U(Z,X)$) and the fact that we are using constrained matchgates become insignificant when dealing with a larger number of qubits. Additionally, here too, we are able to scale up the number of qubits in FermiML models beyond what is feasible with PQC to match the feature count of the dataset.

\section{Conclusion}
FermiML is a QML framework rooted in the physics of free fermions. In this work, we have demonstrated its usefulness as a benchmark of QML approaches on realistic datasets, achieving high levels of accuracy in both binary (WBC) and multi-class (Digits) classification. The latter has seldom been addressed in QML benchmarking~\cite{BAS2024}. In contrast to all other existing QML implementations which are limited by either the capabilities of quantum hardware or the intractability of classical simulation, FermiML is scalable, operating within polynomial runtimes on classical computers.

Our findings prompt further exploration of FermiML models. QML is often advertised as a paradigm that employs quantum resources (entanglement, superposition, interference) to accelerate AI tasks. If FermiML can achieve performance on par with proposed QML methods, then the utility of the latter is questionable. FermiML hence offers a practical tool to benchmark putative QML breakthroughs in realistic settings, thus operationalizing recent skepticism with respect to the applicability of QML as currently construed.~\cite{BAS2024,cerezo2023}. Along this line, it is tempting to benchmark FermiML surrogates of more intricate QML models on larger and more diverse datasets. In fact, one of our ongoing research areas concerning FermiML is its performance across different datasets and various types of data classification tasks.

Another line of investigation is the use of FermiML for classical ``warm-starting'' of QML protocols deployed on actual quantum hardware. This is particularly intriguing given that barren plateaus have so far posed a troublesome obstacle to QML~\cite{MBSBN2018,GWOB2019, ACCCC2021,CSVCC2021,QWZGG2023}. Another promising direction for future research is to explore the use of FermiML for tasks beyond data classification, which would better demonstrate the robustness and utility of FermiML. Finally, FermiML allows for efficiently incorporating quantum mechanical effects in classical deep learning models, along the lines of recent work on quantum-inspired algorithms~\cite{GLT2018,Tang2019,GST2022,ADBL2020,CKLM2023,SBSSBW2023}.

\section*{Acknowledgment}

We acknowledge useful discussions with A.~Chapman and critical reading by S. M. Fadaie.


\appendix

\section{Matchgates and the free-fermionic algorithm}
\subsection{Preliminaries and mapping to free fermions}

Matchgates can be represented as a linear combination of the six operators in the set $\mathcal{M_{G}} =
\{X_{k}X_{k+1}, X_{k}Y_{k+1}, Y_{k}Y_{k+1}, Y_{k}X_{k+1}, Z_{k}I_{k+1}, I_{k}Z_{k+1}\}$. Equivalently, any matchgate acting on nearest-neighbor qubits $k$ and $k+1$ is a Gaussian operation $U = e^{iH}$ generated from a Hamiltonian $H$ represented as a sum of three interactions: 
\begin{align} 
\nonumber
H &= H_1+H_2+H_3, \\ \nonumber
H_{1} &= \alpha_{1} Z_{k}\otimes I_{k+1} + \beta_{1}  I_{k} \otimes Z_{k+1}, \\ \nonumber
H_{2} &= \alpha_{2} X_{k}\otimes X_{k+1} + \beta_{2} Y_{k}\otimes Y_{k+1}, \\
H_{3} &= \alpha_{3} X_{k}\otimes Y_{k+1} + \beta_{3} Y_{k}\otimes X_{k+1},
\end{align}
where $\alpha_{j},\beta_{j}$ are real coefficients and $X_{k}, Y_{k}$ and $Z_{k}$ are Pauli matrices. The Pauli operators are mapped to fermionic creation and annihilation operators via the Jordan-Wigner transformation. Operators related to the j$^{th}$ fermionic mode are expressed in terms of creation ($a_j^{\dagger}$) and annihilation ($a_j$) operators  obeying canonical anti-commutation relations,
\begin{equation}
    \{a_{k}, a_{j}\} = 0,\,\,\{a_{k}^{\dagger}, a_{j}^{\dagger}\} = 0, \,\, \{a_{k}, a_{j}^{\dagger}\} = \delta_{kj}I\,,
\end{equation}
where $j,k = 1,\dots,N$ and $\delta$ denotes the Kronecker delta. With these relations, the Hamiltonian terms can be rewritten as~\cite{Terh2002,MSPMD2023}
\begin{align}
    H_{1} &= 2\alpha_{1} a_{k}^{\dagger}a_{k} +2\beta_{1}a_{k+1}^{\dagger}a_{k+1}, \\
   H_{2} &= \alpha_{2}(a_{k}^{\dagger}-a_{k})(a_{k+1}^{\dagger} +a_{k+1}) -  \beta_{2}(a_{k}^{\dagger}+a_{i})(a_{k+1}^{\dagger}-a_{k+1}), \nonumber\\
    H_{3} &= -i \alpha_{3} (a_{k}^{\dagger}-a_{k}) (a_{k+1}^{\dagger}-a_{k+1}) \nonumber - i \beta_{3} (a_{k}^{\dagger}+a_{k}) (a_{k+1}^{\dagger}+a_{k+1}) \nonumber
\end{align}
and $H$ is then a sum of nearest-neighbor fermionic interactions, quadratic in creation and annihilation operators. Each fermion operator can be split into Majorana operators $c_\mu$ as
\begin{align}
\nonumber
    a_k = \frac{c_{2k-1} + i c_{2k}}{2}, a_k^{\dagger} = \frac{c_{2k-1} - i c_{2k}}{2}\, \\
    c_{2k-1} = a_{k}+a_{k}^{\dagger}, \quad  c_{2k} = -i(a_{k}-a_{k}^{\dagger}) \,,
\end{align}
and the $c_\mu$'s anti-commute: $\{c_{\mu}, c_{\nu}\} = c_{\mu}c_{\nu} + c_{\nu}c_{\mu} = 2 \delta_{\mu \nu} I$, $\,\,\mu,\nu = 1,\dots,2N$. The summation implies that the Majorana monomials are linearly independent, and the associated vector space $\mathcal{C}_{2N}$ has dimension $2^{2N} = 2^N \times 2^N$. Consequently, matrix representations of the $c_{\mu}$'s, are of dimensions, $2^N \times 2^N$.
Invoking the Jordan-Wigner transformation once more, the $2N$-dimensional Majorana operators can be represented in recognised forms of $N$-qubit Pauli operators~\cite{JM08matchgate} as
\begin{align}
\nonumber
c_{2k-1} = Z^{\otimes (k-1)} \otimes X \otimes I ^{\otimes (N-k)}, \\
c_{2k} = Z^{\otimes (k-1)} \otimes Y \otimes I^{\otimes (N-k)}.
\end{align}
The six terms in the matchgate family are then
\begin{align}
\nonumber
Z_{k}I_{k+1} &= -ic_{2k-1}c_{2k}, \\ \nonumber
X_{k}X_{k+1} &= ic_{2k}c_{2k+1},  \\ \nonumber
Y_{k}Y_{k+1} &= ic_{2k-1}c_{2k+2}, \\ \nonumber
Y_{k}X_{k+1} &= ic_{2k-1}c_{2k+1}, \\ \nonumber
X_{k}Y_{k+1} &= -ic_{2k}c_{2k+2}, \\ \nonumber
I_{k}Z_{k +1} &= -ic_{2k+1}c_{2k+2}
\end{align}
for $k = 1,\dots,N$.
\subsection{Simulating output probabilities}
Nearest-neighbor matchgate circuits are recognized for their polynomial-time simulability, attributed to their link with perfect matchings of graphs~\cite{bravyi2008contraction} and their mapping to free-fermionic Hamiltonians. Using this mapping it has been shown that the output probabilities  $|\langle y|U|x \rangle|^{2}$ for any  bitstrings $x$ and $y$ can be evaluated efficiently in polynomial time~\cite{Terh2002,MSPMD2023}. In the following discussion, we recapitulate this method and in the end, provide a detailed explanation tailored to our specific scenario. Our approach leverages matchgates to implement the PQCs, subsequently employed in the execution of the kernel method.
 
To simulate the output probabilities for any input state $|x\rangle$ of length $N$ and Hamming weight $\ell$ it is first expanded into fermionic creation operators acting on the vacuum state (the number of creation operators are determined by the Hamming weight of the bitstring) and in turn are mapped to Majorana operators,
\begin{align}
\nonumber
    |x\rangle &= a_{p_{1}}^{\dagger}...a_{p_{\ell}}^{\dagger} |\boldsymbol{0}\rangle \\ &= c_{2p_{1}}...c_{2p_{\ell}}|\boldsymbol{0}\rangle, \,\,\, \text{with}\,\,\,p_{1}<p_{2}...<p_{\ell}.
\end{align}
The probability that a certain subset $k$ of the total $N$ qubits is in a particular state $y^*$ (meaning, the output bitstring is of length $k$) for a given input state $|x\rangle$ is therefore given by, 
\begin{align}
\label{output1}
    p(y^{*}|x) &= \langle x| c_{2p_{\ell}}...c_{2p_{1}}(U^{\dagger} a_{j_{1}} U)(U^{\dagger} a_{j_{1}}^{\dagger}U) \times \cdots\\ \nonumber & \times(U^{\dagger}a_{j_{k}}^{\dagger} U)(U^{\dagger}a_{j_{k}}U)c_{2p_{1}}...c_{2p_{\ell}}|x\rangle.
\end{align}

Using Eqs.~\eqref{Rmat} and \eqref{Tmat} and the conjugation relations for the fermionic operators
\begin{align}
\label{transfrelns}
    U^{\dagger}a_{j}U &= \sum\limits_{\nu = 1}^{2N} T_{j,\nu}c_{\nu}\,,\,\,\,\\
    U^{\dagger}a_{j}^{\dagger}U &= \sum\limits_{\nu = 1}^{2N} T_{j,\nu}^{*}c_{\nu},
\end{align}
the summation in Eq.~\ref{output1} maybe rewritten as,
\begin{align}
\label{output2}
    p(y^{*}|x) &= \prod\limits_{\gamma = 1}^{k}\sum\limits_{m_{\gamma},n_{\gamma} = 1}^{2N} T_{j_\gamma,m_\gamma}T_{j_\gamma,n_\gamma}^* \\ \nonumber &\times \left\langle \boldsymbol{0} \left|c_{2p_{\ell}}..c_{2p_1}\left(\prod\limits_{\gamma = 1}^{k} c_{m_{\gamma}}c_{n_{\gamma}}\right)c_{2p_1}..c_{2p_{\ell}}
    \right|\boldsymbol{0}\right\rangle. 
\end{align}
For finding the fully contracted terms in the above summation, the Majorana operators are at first expanded in terms of the fermionic creation and annihilation operators. Wick’s theorem of normal ordering of fermions is then applied to each of the terms. All of these fully contracted terms are generated by contraction strategies given in Eqs.(43) and (44) in Ref.~\cite{Terh2002}. The same is summarised using a matrix $B$, $\wick{\c c_{j} \c c_{j}} = B_{ij}$, where $B$ is a  $2N\times 2N$ Hermitian block diagonal matrix, containing $N$ number of $2\times 2$ blocks
\begin{align}
\label{MatB}
    B = \mqty(
        \mqty{
            1 & i \\
            -i & 1 }
        &
        \mqty{
             &  \\
             &  } \\
        \mqty{
             &  \\
             & }
        &
        \mqty{
             \ddots}
        &
        \mqty{
             &  \\
             & } \\
        \mqty{
             &  \\
             & }
        &
        \mqty{
             &  \\
             & }
        &
        \mqty{
            1 & i \\
            -i & 1
        } \\
    ) = \bigoplus\limits_{l = 1}^{N} \mqty(\mqty{
            1 & i \\
            -i & 1 }).
\end{align}

Subsequently Table.~\ref{lookuptab}~\cite{Terh2002} is utilised to construct a $2(\ell+k)\times 2(\ell+k)$ skew-symmetric matrix $M$. The final output probabilities are then simulated using $p(y^*|x) = \Pf(M)=\sqrt{\det(M)}$ where $\Pf$ is the Pfaffian of the matrix. 

\begin{table}[!tbp]
    \centering
    \begin{tabular}{|cc|c|c|c|}
        \hline
        \multicolumn{2}{|c|}{\multirow{2}{*}{$k<j$}} & \multicolumn{3}{|c|}{$j$} \\
        \cline{3-5}
        & & $c_{m_{\beta}}$ & $c_{n_{\beta}}$ & $c_{2p_{\beta}}$ \\
        \hline
        \multirow{4}{*}{$k$} & $c_{m_{\alpha}}$ & $\qty(TBT^\mathsf{T})_{j_{\alpha},j_{\beta}}$ & $\qty(TBT^\dagger)_{j_{\alpha},j_{\beta}}$ & $\qty(TB)_{j_{\alpha},2p_{\beta}}$ \\
        & $c_{n_{\alpha}}$ & $\qty(T^*BT^\mathsf{T})_{j_{\alpha},j_{\beta}}$ & $\qty(T^*BT^\dagger)_{j_{\alpha},j_{\beta}}$ & $\qty(T^*B)_{j_{\alpha},2p_{\beta}}$ \\
        & $c_{2p_{\alpha}}$ & $\qty(BT^\mathsf{T})_{2p_{\alpha},j_{\beta}}$ & $\qty(BT^\dagger)_{2p_{\beta},j_{\beta}}$ & $\delta_{\alpha,\beta}$ \\
        \hline
    \end{tabular}
    \caption[Lookup table]{This lookup table is used to find the matrix elements $M_{kj}$ for $k < j$. $B$ is given in Eq.~\eqref{MatB}.}
    \label{lookuptab}
\end{table}
\subsection{Measurement Scheme for Kernel Method}
Since we employ kernel methods, the output of the circuit is the expectation value of the vacuum state $|\boldsymbol{0}\rangle$, and corresponds to $\langle \boldsymbol{0}|U(x) U^{\dagger}(x')|\boldsymbol{0}\rangle$ where $U(x)$ is the parameterized match-circuit with input data $x$ and $x'$. The kernel is then defined as,
\begin{align}
    \mathcal{U}(x,x') = |\langle \boldsymbol{0} |U^{\dagger}(x')U(x) |\boldsymbol{0}\rangle|^{2} .
\end{align}
and is a similarity measure, portraying the overlap among quantum states related to the data points ($x, x'$). In this specific instance, the Hamming weight remains zero as the expectation value is consistently calculated over the vacuum state and therefore the length of both $y$ and $x$ is $N$. The final matrix $M$ is of dimension $2N$. The Majorana operators originating from the input bitstring  ($c_{2p_{1}},..,c_{2p_{\ell}}$) are therefore absent, leaving only those derived from the conjugation relations in Eqs.\eqref{transfrelns} and \eqref{Rmat}. Therefore Eq.~\eqref{output2} reduces to,
\begin{align}
     p(\boldsymbol{0}|\boldsymbol{0}) = \prod\limits_{\gamma =1}^{N}\sum\limits_{m_{\gamma},n_{\gamma}}^{2N}T_{j_{\gamma},m_{\gamma}}T^{*}_{j_{\gamma},n_{\gamma}}\left\langle \boldsymbol{0} \left|
    \prod\limits_{\gamma =1}^{N}c_{n_{\gamma}}c_{m_{\gamma}}\right|\boldsymbol{0}\right\rangle
\end{align}
This equation is equivalent to the one in Eq.~\eqref{output2} and it is equal to the kernel $\mathcal{U}(x,x')$. 

\begin{figure}[t]
    \centering
    \includegraphics[width=\columnwidth]{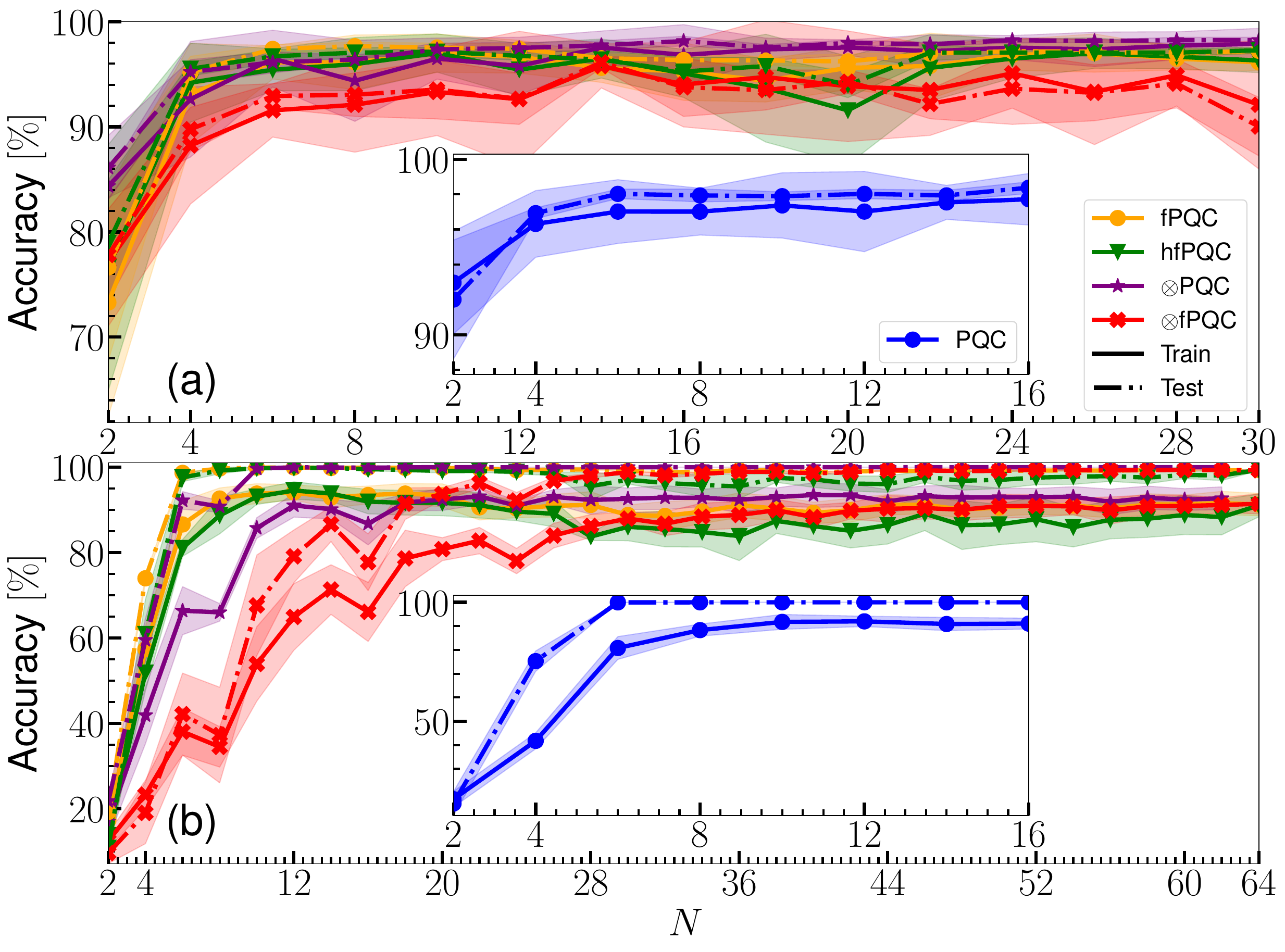}
    \caption{Comparison of training and test accuracies for fPQC, hfPQC, and $\otimes$fPQC for WBC (a) and Digits (b). For each scenario, the insets display a comparative analysis of training and test accuracies for unrestricted PQC and hfPQC. The shaded areas depict the standard deviation of the accuracies.}
    \label{fig:allplotsGA}
\end{figure}

\subsection{Generalization analysis}
In Fig.~\ref{fig:allplotsGA} we provide the training and test accuracies for SVMs optimized with PQC, fPQC, hfPQC, $\otimes$PQC and $\otimes$fPQC kernels for WBC (top row) and Digits (bottom row) datasets. We observe slightly better classification accuracies for the fPQC and hfPQC compared to $\otimes$PQC for smaller qubit counts, up to about $N = 15$. Beyond this point, the test accuracies converge for the DIGITS dataset (Fig.~\ref{fig:allplotsGA}(b)), indicating that the specific details of entangling gates or restriction on the gates become irrelevant when considering larger number of qubits. Upon closer observation, it becomes evident that for smaller qubit counts, the presence of entanglement in the fermiML kernel notably improves classification accuracies. Consequently, $\otimes$fPQC shows significantly worse performance compared to the other models. Additionally, FermiML models show generalization performance (difference between train and test performance) on par with PQCs. Note that the $x$-axes represent the number of qubits in the kernel (and not training iterations as is customary in the ML field).
\end{document}